\documentclass[a4paper,fleqn,usenatbib]{mnras}
\usepackage{newtxtext,newtxmath}
\usepackage[T1]{fontenc}
\usepackage{ae,aecompl}
\usepackage{epsfig}
\usepackage{amsmath, amsfonts, epsfig, xspace}
\usepackage{natbib}
\usepackage{deluxetable}
\usepackage{rotating}
\usepackage{graphicx}
\usepackage{caption}
\usepackage{longtable}
\usepackage{multicol}
\usepackage{booktabs}
\usepackage{dcolumn}
\usepackage{longtable}
\usepackage[labelfont=bf]{caption}
\usepackage{subfig}
\usepackage{floatrow}
\usepackage{float}
\usepackage[morefloats=200]{morefloats}
\usepackage{threeparttablex}
\def\aj{{AJ}}%
\def\araa{{ARA\&A}}%
\def\apj{{ApJ}}%
\def\apjl{{ApJ}}%
\def\apjs{{ApJS}}%
\def\aap{{A\&A}}%
\def\aapr{{A\&A~Rev.}}%
\def\aaps{{A\&AS}}%
\def\mnras{{MNRAS}}%
\def\pasp{{PASP}}%
\title[INOV of low-mass AGN]{Intranight optical variability of low-mass Active Galactic Nuclei: A Pointer to blazar-like activity}
\author[$Gopal-Krishna$ et al.]{{ \large Gopal-Krishna$^{1}$,
Krishan Chand$^{2,3}$, Hum Chand$^{4}$\thanks{E-mail: humchand@gmail.com (HC)}, Vibhore Negi$^{2}$, Sapna Mishra$^{5}$, S. Britzen$^6$, and P. S. Bisht${^7}$}\\\\
$^{1}$UM-DAE Centre for Excellence in Basic Sciences, Vidyanagari, Mumbai 400098, India\\
$^{2}$Aryabhatta Research Institute of Observational Sciences (ARIES), Manora Peak, Nainital 263002, India\\
$^{3}$Department of Physics, Kumaun University, Nainital 263002, India\\
$^{4}$Department of Physics and Astronomical Science, Central University of Himachal Pradesh (CUHP), Dharamshala 176215, India\\
$^{5}$Inter-University Centre for Astronomy and Astrophysics (IUCAA), Postbag 4, Ganeshkhind, Pune 411007, India\\
$^{6}$Max-Planck-Institut f. Radioastronomie, Auf den Huegel 69, 53121 Bonn, Germany \\
  $^{7}$Department of Physics, Soban Singh Jeena University, Almora 263601, India\\}
\begin{document}
\date{Accepted ---; Received ---; in original form ---}

\pagerange{\pageref{firstpage}--\pageref{lastpage}} \pubyear{2022}

\maketitle

\label{firstpage}

\begin{abstract}
This study aims to characterise, for the first time, intranight optical variability (INOV) of low-mass active galactic nuclei (LMAGN) which host a black hole (BH) of 
mass $M_{BH} \sim 10^6 M_{\odot}$, i.e., even less massive than the Galactic centre black hole Sgr A* and 2-3 orders of magnitude below the supermassive black holes (SMBH, $M_{BH}$ $\sim$ $10^8 - 10^9 M_{\odot}$) which are believed to power quasars. Thus, LMAGN are a crucial subclass of AGN filling the wide gap between SMBH and stellar-mass BHs of Galactic X-ray binaries. We have carried out a 36-session campaign of intranight optical monitoring of a well-defined, representative sample of 12 LMAGNs already detected in X-ray and radio bands. This set of LMAGN is found to exhibit INOV at a level statistically comparable to that observed for blazars (M$_{BH} \gtrsim$ 10$^{8-9}$ M$_{\odot}$) and for the $\gamma$-ray detected Narrow-line Seyfert1 galaxies (M$_{BH}\sim 10^7$ M$_{\odot}$) which, too, are believed to have relativistic jets. This indicates that the blazar-level activity can even be sustained by central engines with black holes near the upper limit for Intermediate Mass Black Holes ($M_{BH}$ $\sim$ $10^3 - 10^6 M_{\odot}$).
\end{abstract}
\section{Introduction}
\label{introduction}

Luminous Active Galactic Nuclei (AGN) of massive galaxies are believed to be powered by accretion onto supermassive black holes (SMBH) of masses $M_{BH}$ $\gtrsim$ $10^{8} M_{\odot}$. Powerful nonthermal radio jets are known to be ejected by a significant minority of such SMBH and also by accreting stellar-mass BHs in the Galaxy (cf. reviews by \citealt{Urry1995PASP..107..803U,2019ARA&A..57..467B}). However, evidence is very sparse about such activity in BHs that fill the huge mass gap between supermassive and stellar-mass BHs, particularly about `Intermediate Mass Black Holes' (IMBH) with $M_{BH}$ $\sim$ $10^3 - 10^6 M_{\odot}$ (e.g., \citealt{2020ARA&A..58..257G} and references therein; \citealt{Seepaul2022MNRAS.515.2110S}).
Such BHs are thought to exist in moderately massive, or dwarf galaxies (e.g., \citealt{2013ARA&A..51..511K}; \citealt{2016ASSL..418..263G}). 
{Although, essentially no information is presently available about the jet-forming capability of such black holes, work over} the past $\sim$ 25 years has found evidence for `normal' AGN activity occurring in galaxy centers harboring BHs of masses down to $\sim$ $10^{5} M_{\odot}$ (\citealt{2020ARA&A..58..257G}); with those  having $M_{BH}$ above $\sim$ $10^{6.5}$ $M_{\odot}$ typically deemed to be normal AGN (e.g., \citealp{Qian2018ApJ...860..134Q}).\par
\vspace{-1mm}
A generic { manifestation of AGN activity is flux variability across the spectrum}, most conspicuously observed in the tiny subset, called blazars. These include BL Lac objects (BLLs) and flat-spectrum radio quasars (FSRQs), specifically their high-polarization subset, called HPQs.
Intensity of blazars is usually dominated by relativistic jets of nonthermal radiation Doppler boosted in our direction. Their observed strong variability is thought to arise from shocks or bulk injection of energetic particles in the jet  (e.g., \citealt{1985ApJ...298..114M}; \citealt{Valtaoja1992A&A...254...80V}; \citealt{Spada2001MNRAS.325.1559S}; \citealt{Giannios2010MNRAS.408L..46G}), or, alternatively, due to jet helicity/swings (\citealt{Camenzind1992A&A...255...59C}; \citealt{Gopal-krishna1992A&A...259..109G}), or jet precession (e.g., \citealt{2000A&A...355..915A}; \citealt{2018MNRAS.478.3199B}). 
{The rapid (hour/minute-like) flux variability of blazars
has been most conspicuously detected at TeV energies, on time scales as short as a few minutes \citep{Aharonian2007ApJ...664L..71A,Albert2007ApJ...669..862A}, which are much shorter than light-crossing times at a $10^8 M_{\odot}$ black hole's  horizon ($\sim$ 15 minutes), 
suggesting that the variability involves small regions of enhanced emission within an outflowing jet \citep{Begelman2008MNRAS.384L..19B}.} Rapid variability has been most extensively documented in the optical band and termed `Intra-Night Optical Variability' (INOV, \citealt{2003ApJ...586L..25G}). {A likely cause of INOV of blazars, too, is a strong relativistic enhancement of small fluctuations arising through turbulent pockets in the jet \citep[][]{Marscher1991STIA...9352913M,2012A&A...544A..37G,Calafut2015JApA...36..255C}.
The occurrence of strong INOV of HPQs (blazars which are always radio-loud) and the low-polarization radio quasars (LPRQs) was compared by \citet{2012A&A...544A..37G}, by carrying out sensitive and densely sampled intranight optical monitoring of 9 HPQs and 12 LPRQs (both having flat radio spectrum). Remarkably, the HPQ subset showed strong INOV (i.e., amplitude  $\psi > 4$\%) on 11 out of 29 nights, in stark contrast to the LPRQs for which strong INOV was observed on just 1 out of 44 nights. Evidently, a strong INOV can be an effective tracer of blazar activity in AGN. Here, it may be mentioned that (low-level) INOV has also been observed in radio-quiet quasars (RQQs) which launch at most feeble jets (e.g., \citealp{2016ApJ...831..168K}). Their INOV may well be associated with (transient) shocks (\citealp{Chakrabarti1993ApJ...411..602C}) and/or `hot spots' (\citealp{Mangalam1993ApJ...406..420M}) in the accretion disk around the SMBH. However, INOV of RQQs is almost always weak ($\psi < 3\%$) and even that is detected in just $\sim 10\%$ of the sessions, which is similar to the 
INOV pattern observed for non-blazar type jetted AGN, such as lobe-dominated quasars and even weakly polarised radio core-dominated quasars (\citealp{Stalin2004JApA...25....1S}; \citealp{Goyal2013MNRAS.435.1300G}; \citealp{2018BSRSL..87..281G}).

A related class of AGN exhibiting strong INOV with a fairly high duty cycle of $\sim$ 25\% (\citealt{2021MNRAS.501.4110O}; also, \citealt{Paliya2013MNRAS.428.2450P}) is the
tiny, radio-loud minority of Narrow-line Seyfert1 (NLS1) galaxies which has extreme properties similar to blazars, like a flat radio spectrum, high brightness temperature (e.g., \citealp{2008ApJ...685..801Y}) and $\gamma-ray$ detection, all of which are widely interpreted in terms of a relativistically beamed jet (e.g., \citealp{Abdo2009ApJ...699..976A}; \citealt{Foschini2011nlsg.confE..24F};  \citealt{2019ARA&A..57..467B}). This premise is further confirmed by the direct detection of VLBI jets in such sources \citep [e.g.,][] {Giroletti2011A&A...528L..11G}. While all this evidence affirms the relativistic jet connection to their strong INOV, an important difference from blazars is that the BHs in radio-loud NLS1 galaxies are estimated to be typically an order-of-magnitude less massive ($\sim$ $10^{7} M_{\odot}$; e.g., \citealt{2008ApJ...685..801Y}; \citealt{2012ApJ...755..167D}; \citealt{2020Univ....6..136F}), estimated through application of the virial method (e.g., \citealt{2000ApJ...533..631K}; \citealt{1999ApJ...526..579W}; \citealt{2006ApJ...641..689V}) which employs FWHM of the broad-line region (BLR) emission lines and the continuum luminosity measured by single-epoch optical spectroscopy. Thus, powerful relativistic jets do appear to get ejected even by moderately massive BHs, although indications are that the bulk Lorenz factor of such jets is typically smaller with respect to blazar jets (\citealt{2019ApJ...872..169P}), an inference also supported by radio observations (\citealt{2015A&A...575A..55A}; \citealt{2015ApJS..221....3G}; \citealt{2016RAA....16..176F}). 

Coming to a further one order-of-magnitude less massive BHs powering LMAGNs, the question arises if they are at all capable of blazar-like activity? This question is addressed here by determining their INOV properties which, as mentioned above, can be an effective discriminator between blazars and non-blazars. In this first such attempt we have used a representative sample of 12 LMAGN whose central engines have masses close to 
$10^6 M_{\odot}$, i.e., between the lower mass end for normal AGN and the upper mass end of IMBH. These LMAGN are less massive than even the black hole Sgr A* located at the nucleus of our Galaxy ($M_{BH}$ $\sim$ 4 $\times 10^6 M_{\odot}$ \citealt{2018A&A...615L..15G}; \citealt{2021ApJ...917...73W}), in which a faint nonthermal radio jet has been detected \citep{Yusef2020MNRAS.499.3909Y}. In an extreme event, this highly variable source has shown a factor of 75 change in flux over a 2 hr time span, 
at near-infrared wavelengths where, unlike the optical band, its emission is not masked out by the opacity of the intervening galactic material (see, \citealt{Do2019ApJ...882L..27D}).}  
\vspace{-5.0mm}
\section{The sample of low-mass AGN (LMAGN)}
\label{The IMBH sample}
Our sample of 12 LMAGN has been extracted from a well-defined set of 29 such broad-line objects, assembled by \citet{Qian2018ApJ...860..134Q}, using the criteria of a detected counterpart in X-ray and radio bands, and a $M_{BH}$ in the range $\sim$ $10^{5.5} - 10^{6.5} M_{\odot}$ (These objects were selected by them from \citealt{2012ApJ...755..167D} and other literature). Their BH masses had been determined by applying the viral estimator to carefully measured width and luminosity of the broad 
$H_{\alpha}$ line in single-epoch spectrum and, according to \citet{Qian2018ApJ...860..134Q} the typical uncertainty of these mass estimates is $\sim$ 0.6 dex. To these 29 LMAGN we applied a cut in the g-band (SDSS) magnitude, $m_g$ < 17.0 (for two sources, J010927.1+354305 and J083615.12-262434.16, lacking  {\it $m_g$} (SDSS), we used $m_V$ (SIMBAD) together with the transformation equation of \citealp{Jester2005AJ....130..873J}). 
The shortlisted 13 LMAGN had one source (J083615.12-262434.16) located  
too far south for our telescopes, and its exclusion led to our final sample of 12 LMAGN (Table \ref{tab1}). Their $M_{BH}$ values are tightly clustered around the median of $10^6 M_{\odot}$, and none exceeds $2\times10^6 M_{\odot}$. As seen from Table \ref{tab1}, all these AGN have z $\lesssim$ 0.1 and a radio-loudness parameter 
$R_{5 GHz}$ < 10 ($R_{5GHz}$ is the ratio of radio to optical flux densities, \citealp{Kellermann1989AJ.....98.1195K}). This value is well within the conventional upper limit of $R_{1.4GHz}$ = 19 for radio-quiet quasars, when translated to 1.4 GHz and B-band (e.g., \citealt{2008ApJ...685..801Y}; \citealt{Komossa2018rnls.confE..15K}).
Their `radio-quiet' classification is also consistent with the radio luminosity limit of $5\times10^{29}$ erg Hz$^{-1}$ at 1.4 GHz specified for radio-quiet quasars (RQQs), since such radio luminosities can even arise from star formation in the host galaxy (\citealt{2016ApJ...831..168K}). 
\floatsetup[table]{capposition=top}
\begin{table*}
  \centering
\begin{minipage}{180mm}
    \caption{Basic properties of the sample of 12 X-ray and radio detected `Low-mass AGN' (LMAGN).}
    \label{tab1}
    \resizebox{\textwidth}{!}{%
\begin{tabular}{cccccccccccc}
\hline\\
\multicolumn{1}{c}{Source} &\multicolumn{1}{c}{$z$} &\multicolumn{1}{c}{ $m_{g}$}  &\multicolumn{1}{c}{$m_B$} & $M_B$    &\multicolumn{1}{c}{Galaxy} &\multicolumn{1}{c}{log${M}_{{BH}}$}&\multicolumn{1}{c}{log$({L}_{{b}}/{L}_{{Edd}})$} &Flux at& \multicolumn{1}{c}{Flux at} &Radio &$P_{1.4\,(GHz)}$  \\
 SDSS name                 &                        &                           &              &         &type                        &                                          &                                                                  &B      &1.4 GHz                       &loudness                         &                \\
                          &  NED                      &    SDSS                              &              &         &   SDSS                    &                                          &                                                                  & (mJy)     &  (mJy)                     & $R_{1.4(GHz)}$     & (erg$s^{-1}Hz^{-1}$)                    \\ 
(1)                        &   (2)                   &(3)                   &(4)                      & (5)     &(6)                     &(7)                    &(8)      & (9)                     &(10)                        &(11)      &(12)               \\\\
 \hline\\
J010927.02+354305.00 &      0.00006$^a$       &12.27$^b$&12.56   &$-$15.60     & SA0$^{-(s)}$$^{1}$ &5.65&$-$5.5$\dagger$&42.85   &    3.40* &0.08 & 7.48 $\times$ $10^{25}$\\     
J030417.70+002827.40 &     0.04443         &15.83         &17.86   &$-$18.50     & Sc$^{2}$       &6.20&$-$0.6         &  0.33   &    0.62 	 &1.88 & 2.60 $\times$ $10^{28}$\\    
J073106.87+392644.70 &     0.04832         &15.89         &19.07   &$-$17.50     & Sbc$^{2}$      &6.00&$-$0.7         &  0.11   &    0.61 	 &5.55 & 3.10 $\times$ $10^{28}$\\    
J082443.29+295923.60 &     0.02542         &15.92         &16.32   &$-$18.80     & $S0_{-}^{2}$   &5.70&$-$0.8*        &  1.34   &    1.67 	         &1.25 & 2.23 $\times$ $10^{28}$  \\
J082433.33+380013.10 &     0.10316         &16.63         &16.81   &$-$21.50     &Spiral*         &6.10&$-$0.4         &   0.86   &    1.07 	 &1.24 & 2.70 $\times$ $10^{29}$\\
J085152.63+522833.00 &      0.06449        &16.45         &19.16   &$-$18.10     &Uncertain*      &5.80&$-$0.6         &   0.10   &    0.92 	 &9.20 & 8.83 $\times$ $10^{28}$\\
J104504.23+114508.78 &      0.05480        &16.56         &16.99   &$-$19.90     &Uncertain*      &6.20&$-$0.8         &   0.72   &    0.82 	 &1.14 & 5.60 $\times$ $10^{28}$\\
J110501.99+594103.70 &      0.03369        &15.15         &15.07   &$-$20.70     &Spiral*         &5.58&$-$0.5*        &   4.25   &    5.96$\dagger$ &1.40 & 1.45 $\times$ $10^{29}$\\    
J122548.86+333248.90 &     0.00106         &14.24         &10.80   &$-$17.30     & SA(s)m0$^{1}$   &5.56&$-$2.9$\dagger$&  216.77 &    1.12 	 &0.01& 2.33 $\times$ $10^{25}$ \\
J132428.24+044629.70 &     0.02133         &16.09         &16.46$^c$  &$-$18.51$^e$   &S0/a$^{2}$      &5.81&$-$1.4         &  1.18   &    1.79 	 &1.52 & 2.09 $\times$ $10^{28}$ \\
J140040.57$-$015518.30&    0.02505       &15.91         &17.00   &$-$18.10     & compact$^{3}$ &6.30&$-$1.4           &  0.72   &    1.75 	 &2.43 & 2.30 $\times$ $10^{28}$  \\  
J155909.63+350147.50 &      0.03148        &14.61         &15.69   &$-$20.00     & SB(r)b$^{1}$    &6.31&$-$0.2*        &  2.40   &    2.72 	 &1.13 & 6.15 $\times$ $10^{28}$ \\\\
\hline
\multicolumn{12}{l}{{\bf Col. 2}: `$a$': \citet{Wilson2012MNRAS.424.3050W}.; {\bf Col. 3}: m$_g$ from SDSS DR14 \citep{Abolfathi2018ApJS..235...42A} and the g-mag marked 
`$b$' is estimated from $m_B$ \& $m_V$ (SIMBAD) using the }\\
 \multicolumn{12}{l}{transformation given by \citet{Jester2005AJ....130..873J}; {\bf Col. 4}: From \citet{Veron2010AA...518A..10V}. The entry marked `$c$' has been estimated from $m_u$ \& $m_g$ (SIMBAD)
 using the }\\
 \multicolumn{12}{l}{transformation given by \citet{Jester2005AJ....130..873J}; {\bf Col. 5}: $M_B$ taken from \citet{Veron2010AA...518A..10V}. `$e$' Estimated from $m_B$, using the distance modulus taken from NED;   }\\
  \multicolumn{12}{l}{{\bf Col. 6}: `*' SDSS DR14 \citep{Abolfathi2018ApJS..235...42A}, (1): \citet{de1991rc3..book.....D}, (2): \citet{Nair2010ApJS..186..427N}, (3): \citet{Zwicky1975AJ.....80..545Z}; {\bf Col. 7}: Taken from \citet{Qian2018ApJ...860..134Q};  }\\
   \multicolumn{12}{l}{{\bf Col. 8}: Eddington ratio, taken from \citet{2012ApJ...755..167D} except for those marked with `*' and `$\dagger$', for which the references are \citet{Greene2007ApJ...670...92G} and \citet{Nyland2012ApJ...753..103N}, }\\
  \multicolumn{12}{l}{respectively; {\bf Col. 9}: Calculated from $m_B$, following \citet{Schmidt1983ApJ...269..352S}; {\bf Col. 10}: Taken from \citet{Qian2018ApJ...860..134Q}, except for the sources marked with `$\dagger$' (FIRST survey,  }\\
    \multicolumn{12}{l}{\citealp{Becker1995ApJ...450..559B}, \citealp{White1997ApJ...475..479W}) and `*' (NVSS, \citealt{Condon1998AJ....115.1693C}); {\bf Col. 11}: Radio-loudness parameter $R_{1.4\,(GHz)} = $$flux_{1.4\,(GHz)}$ / $flux_{B}$; {\bf Col. 12}: Luminosity at   }\\
   \multicolumn{12}{l}{1.4 GHz, calculated from $flux_{1.4}\,(GHz)$  (col. 10) and distance d estimated from the distance modulus, i.e., $m_B\,-\,M_B\,=\,5log\,(d)\,-\,5$.}\\
  \end{tabular}
 }
 \end{minipage}
  \end{table*}  
\section{Observations, data reduction and analysis} \label{Observations and analysis}
Three sessions of minimum 3 hour duration each were devoted to each of the 12 sources (online Table S2). The monitoring was done in the R-band (where the CCD detector used has maximum sensitivity), using the 1.3-metre Devasthal Fast Optical Telescope  (DFOT; \citealt{Sagar2011CSci..101.1020S}, 34 sessions) and the 1.04-metre Sampuranand Telescope (ST; \citealt{Sagar1999CSci...77..643S}, 2 sessions), both operated by the Aryabhatta Research Institute of observational sciencES (ARIES), Nainital, India. Details of the observational set-up and procedure can be found in \citet{Sapna2019MNRAS.489L..42M} and \citet{2020MNRAS.493.3642O}. A log of the observations, together with data on the comparison stars used here for differential photometry, are provided in the online Table S1. 
\begin{figure*}
 \includegraphics[width=0.9\textwidth,height=0.28\textheight]{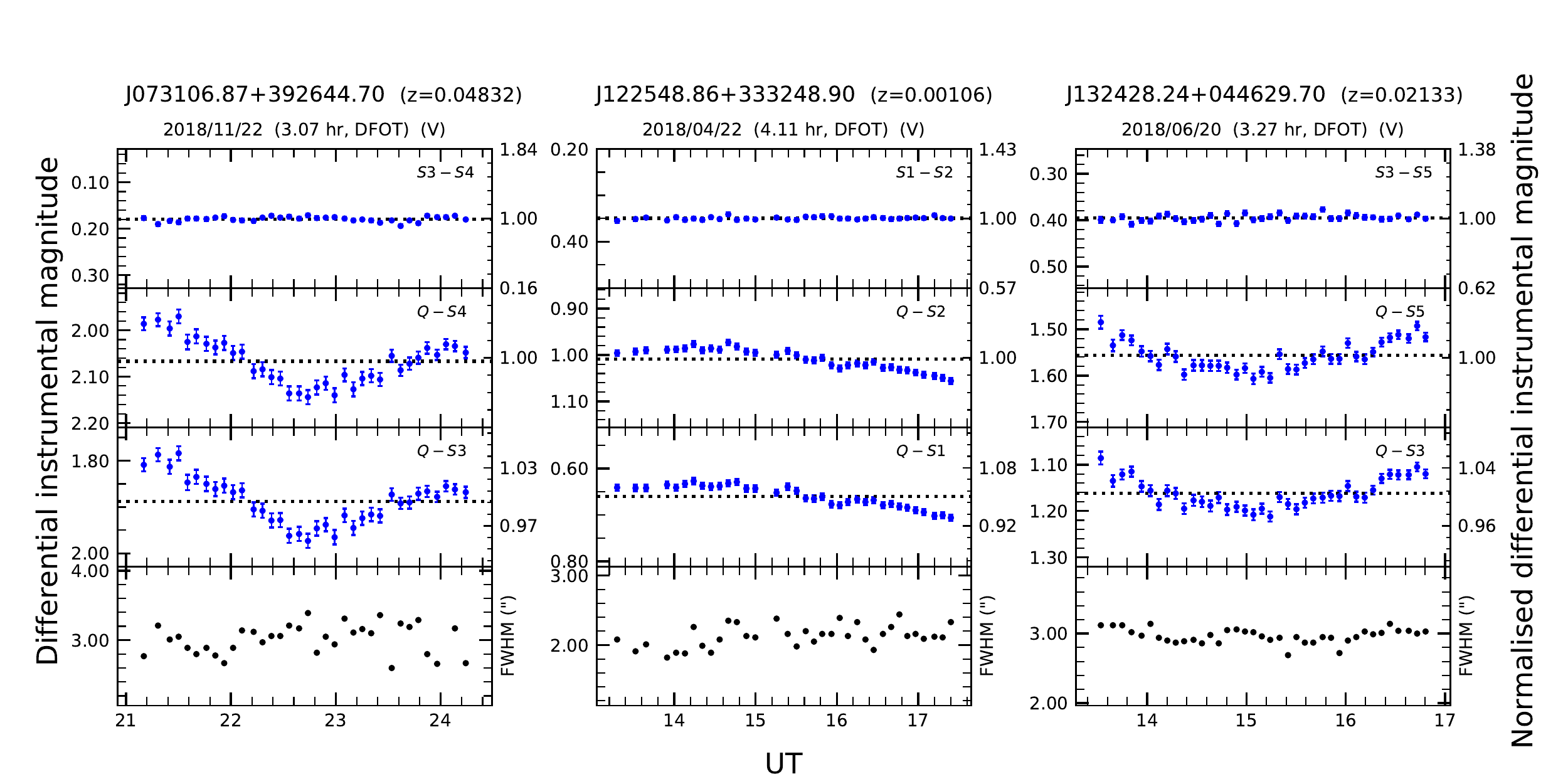}\\
 \vspace{-0.1in}
  \caption{These plots for 3 out of the total 12 monitoring sessions with INOV detection show differential light curves (DLCs) of the AGN relative to two comparison stars, as well as the star1-star2 DLC (top panel), as well as the run of the `point spread function' (PSF, see the bottom panel for each session). The labels on the right side of each panel show the  differential instrumental magnitude normalised by  mean value, as marked by the dotted horizontal line.}
\label{fig1} 
\end{figure*}

Pre-processing of the raw images (bias subtraction, flat-fielding and cosmic-ray removal) was done using the standard tasks available in the Image Reduction and
Analysis Facility (IRAF)\footnote{\url{http://iraf.noao.edu/}}. {The instrumental magnitudes of the target AGN and the chosen (non-varying) comparison stars in the same CCD frame were determined for each frame, by aperture
photometry \citep[]{Stetson1987PASP...99..191S,1992ASPC...25..297S}, using the Dominion Astronomical Observatory Photometry II (DAOPHOT II algorithm). Also, for each
frame, we determined the `point spread function' (PSF) by averaging the FWHMs of the profiles of 5 bright unsaturated stars within the frame. Median of the PSF values for
all the frames in a session gave the `seeing' (FWHM of the PSF) for the session. Based on prior experience, we took 2$\times$FWHM as the aperture radius for photometry for the session. The estimated INOV parameters for the sample were found to remain essentially unchanged when, as a check, aperture radius was set equal to 3$\times$FWHM.
Since our targets are nearby AGN ({\it z} $\lesssim$ 0.1), their aperture photometry may have a significant contribution from the underlying host galaxy.} It is therefore important that the PSF does not have any systematic drift through the session (see, \citealt{Cellone2000AJ....119.1534C}; \citealt{Nilsson2007A&A...475..199N}). This is indeed the case for a large majority of our 36 sessions, as seen from the bottom panel for each session (online Figs. S1-S4). If, however, a PSF gradient was found for a session, no claim of INOV detection was made for that session. We followed this conservative approach in order to ensure that the INOV detections claimed here for our sample of intrinsically weak nearby AGN are not an artefact due to a gradient in PSF.  Figs. S1-S4 of the on-line material show for each session, the differential light curves (DLCs) of the AGN, relative to two comparably bright steady comparison stars, as well as the `star1 - star2' DLC and the run of PSF through the session. Results for 3 of the sessions are displayed in Fig. \ref{fig1}. To check for INOV in a session, both DLCs of the AGN were subjected to the F-test (\citealt{Diego2010AJ....139.1269D}; \citealt{Goyal2013MNRAS.435.1300G}). Details of the procedure are provided in the online material (Appendix A), together with the methodology for computing the AGN's fractional variability amplitude ($\psi$) and the duty cycle (DC) of INOV for the sample. The duty cycle was computed according to the following definition \citep{Romero1999A&AS..135..477R}:
\begin{equation} 
DC = 100\frac{\sum_\mathbf{i=1}^\mathbf{n} N_i(1/\Delta T_i)}{\sum_\mathbf{i=1}^\mathbf{n}(1/\Delta T_i)} {\rm \%} 
\label{eq:dc} 
\end{equation}
 Here $\Delta T_i = \Delta T_{i,obs}(1+z)^{-1}$ is the intrinsic duration of the $i^{th}$ session, obtained by correcting for $z$ of the source ($\Delta T_{i,obs}$ is 
 listed in column 4 of the online Table S2).
 If variability was detected in the $i^{th}$ session, $N_i$ was taken as 1, otherwise $N_i$ = 0.

Additional information on the methodology of analysis can be found in \citet{Goyal2013MNRAS.435.1300G} and \citet{Krishan2022MNRAS.511L..13C}. We reiterate that most of the 36 sessions did not witness a systematic drift in PSF (online Figs. S1-S4) and all INOV detections claimed here are limited to such sessions only. 
\vspace{-5.0mm}
\section{\bf Results and discussion}

This study presents the first characterisation of INOV for low-mass AGN (LMAGN) whose black holes have masses tightly clustered around the median value of $10^6 M_{\odot}$ and all of which have been detected in the X-ray and radio bands. A well-defined representative sample of 12 such LMAGN (Table \ref{tab1}) was monitored by us in 36 sessions and the resulting DLCs were examined for INOV by applying the F-test (sect. \ref{Observations and analysis}). Significant INOV was detected in 8 sessions (online Table S2) and a mosaic of 3 of these type `V' sessions is displayed in Fig. \ref{fig1}. Another two sessions were placed under `probable variable' (PV) category. All type `V' sessions showed an INOV amplitude $\psi > 3\%$ and the corresponding duty cycle (DC) of INOV is found to be $\sim$ 22\% ($\sim$ 28\%, if the two `PV' type sessions are also included). Here we recall the extensive INOV study published by \citet{Goyal2013MNRAS.435.1300G}, who followed a very similar analysis procedure and covered 6 prominent classes of powerful AGN monitored in 262 sessions, showed that an INOV amplitude $\psi > 3-4\%$ and a duty cycle  above $\sim$ 10\% (for $\psi > 3\%$) are only observed for blazars (a DC ($\psi > 3\%$) $\sim$ 35\% was found by them for blazars). A comparison of these values with the present results reveals a blazar-like level of INOV for the present sample of LMAGNs. It may also be noted that the present estimate of INOV DC ($\psi > 3\%$) $\sim$ 22\% for LMAGNs is likely to be an underestimate because the aperture-photometry of these nearby, low-luminosity,  AGN is likely to be significantly contaminated by `non-variable' emission contributed by the host galaxy, diluting the variable component (e.g., see \citealt{2021MNRAS.501.4110O}). Another potential cause of DC underestimation, as mentioned above, is our conservative approach of restricting claims of INOV to only those sessions during which `seeing' remained steady, i.e., the PSF showed no systematic gradient (online Figs. S1-S4).

Whilst the detection of blazar-like INOV levels among the LMAGN is a pointer to the presence of relativistically beamed jets in them, it is a somewhat unexpected result, on two counts. Firstly, observations indicate that, in comparison to (SMBH powered) blazars, radio-loud NLS1s (including their $\gamma$-ray detected subset)  powered by an order-of-magnitude less massive BHs, are prone to having only mildly relativistic jets (\citealt{2015ApJS..221....3G}; \citealt{2015A&A...575A..55A}). 
Extrapolating this trend to even lower $M_{BH}$ range being probed here, one would expect their jets to be at most mildly relativistic and therefore exhibit only low-level INOV. Secondly, all the 12 LMAGNs monitored here formally belong to the radio-quiet category (sect. \ref{The IMBH sample}), according to both conventionally adopted criteria, namely radio luminosity and the radio-loudness parameter (R) (see, however, \citealt{2001ApJ...555..650H}). Since powerful radio-quiet quasars energised by SMBH, almost never exhibit a strong INOV with $\psi$ > 3-4\% (e.g., \citealt{2018BSRSL..87..281G} and references therein), the strong INOV levels found here for the LMAGN appear striking. However, it may be reiterated that these LMAGN are not radio-silent and their low radio luminosities are consistent with the non-linear dependence of jet power on $M_{BH}$ ($P_{jet}$ $\propto$ $M_{BH}^{{17/12}}$, \citealt{Heinz2003MNRAS.343L..59H}; see also, \citealt{2003MNRAS.340.1095D}). Also, their INOV behaviour may be viewed from the perspective of the recent detection of flaring events at 37 GHz in `radio-silent' NLS1s, occurring several times a year, which has strengthened the case for the capacity of even such radio-silent AGN to launch relativistic jets (\citealt{Lahteen2018A&A...614L...1L}). This demonstrates that, at least for moderately massive AGN, the observed large variability of millimetric flux, probably arising from a relativistic jet, is essentially decoupled from their `radio-quietness'. It seems reasonable to expect that the optical emission is more closely tied to nuclear jet emission at millimetre wavelengths as compared to its emission at lower frequencies (i.e., radio), which is prone to heavy attenuation due to high opacity on the innermost sub-parsec scales of the nuclear jet, as inferrred from VLBI studies (e.g., \citealp{Boccardi2017A&ARv..25....4B}; \citealt{Gopal-krishna1991vagn.conf..194G}).
Conceivably, an extension of the above trend to even less massive AGN (LMAGN) being probed here, could then explain the strong INOV activity observed in them, despite their being formally `radio-quiet' (albeit detected in both X-ray and radio bands). Here, it is interesting to recall that for NLS1s, \citet{Lister2018rnls.confE..22L} have underscored the importance of indicators other than radio-loudness, such as flux variability, as proof of relativistic jet.

The present evidence for jet activity in LMAGN consolidates their usefulness to studies of the `Fundamental Plane' (FP) of BH activity. This powerful tool,
discovered by \citet{Merloni2003MNRAS.345.1057M} and \citet{Falcke2004A&A...414..895F} has played a prominent role in unifying supermassive black holes powering AGN, with solar-mass galactic black holes (GBH) powering X-ray emitting stellar binaries (see, also, \citealt{Falcke1996A&A...308..321F}; \citealt{Heinz2003MNRAS.343L..59H}; \citealt{Fender2004MNRAS.355.1105F}). Basically, FP relates mass accretion rate, proxied by X-ray luminosity, to the jet or outflow power, probed by the radio luminosity, at a given $M_{BH}$. It is meant to extend the hard-state GBH radio/X-ray correlation (\citealt{Gallo2003MNRAS.344...60G}; \citealt{Corbel2003A&A...400.1007C}; \citealt{Maccarone2003MNRAS.345L..19M}) to include a mass term and thus link the hard-state GBH and their supermassive analogues, altogether spanning $\sim$ 8 orders-of-magnitude in $M_{BH}$. With the evidence for blazar activity, presented here, LMAGN ($M_{BH}$ $\sim$ $10^6 M_{\odot}$) can play a crucial role in defining the FP at the intermediate mass range. Being 2-3 magnitudes less massive than the SMBH powering blazars, the X-ray segment of their broadband SEDs would be much less affected by synchrotron cooling, further aided by the spectral Doppler shift pushing the SED to higher energies (e.g., \citealt{Plotkin2016ApJ...825..139P}; also, \citealt{Gultekin2019ApJ...871...80G}). These two factors would make the observed X-ray emission of LMAGN provide a more reliable measure of the mass accretion rate, leading to a more precise determination of the Fundamental Plane.
\vspace{-8.0mm}
\section{Conclusions}
We have presented the first attempt to characterise intranight optical variability of low-mass AGN (LMAGN) whose black holes have masses tightly clustered around $10^6 M_{\odot}$, i.e., near the low-mass end for AGN and even lower than the mass of the BH at the centre of our Galaxy. For this, we monitored in 36 sessions a well-defined representative sample of 12 LMAGNs, each with detected X-ray and radio emission, albeit formally in the radio-quiet domain.  The specific question we have addressed is whether such LMAGN can support nuclear activity, like blazars which are powered by BHs 2-3 orders-of-magnitudes more massive.
The observed similarity of their INOV level to that of blazars hints at the presence of relativistic jets in our set of LMAGNs, akin to the inference reached for `radio-silent' NLS1 galaxies from their observed flaring at millimetre wavelengths (\citealt{Lahteen2018A&A...614L...1L}). It is further noted that such LMAGN can play an important role in reliable determination of the `Fundamental Plane' of jet activity in black holes.

\vspace{-7.5mm}

\section*{Acknowledgments}
We are thankful to the anonymous referee for constructive suggestions. G-K acknowledges a Senior Scientist fellowship of the Indian National Science Academy. The assistance from the scientific and technical staff of ARIES DFOT and ST is thankfully acknowledged. 
\vspace{-5.5mm}
\section*{Data availability}
The data used in this study will be shared on reasonable request to the corresponding author.
\vspace{-6.5mm}

\bibliography{references}
\onecolumn
\renewcommand{\thetable}{S\arabic{table}}
 \addtocounter{table}{-1}
 \renewcommand{\thefigure}{S\arabic{figure}}
  \addtocounter{figure}{-1}

\begin{figure*}
 \includegraphics[width=1.09\textwidth,height=0.95\textheight]{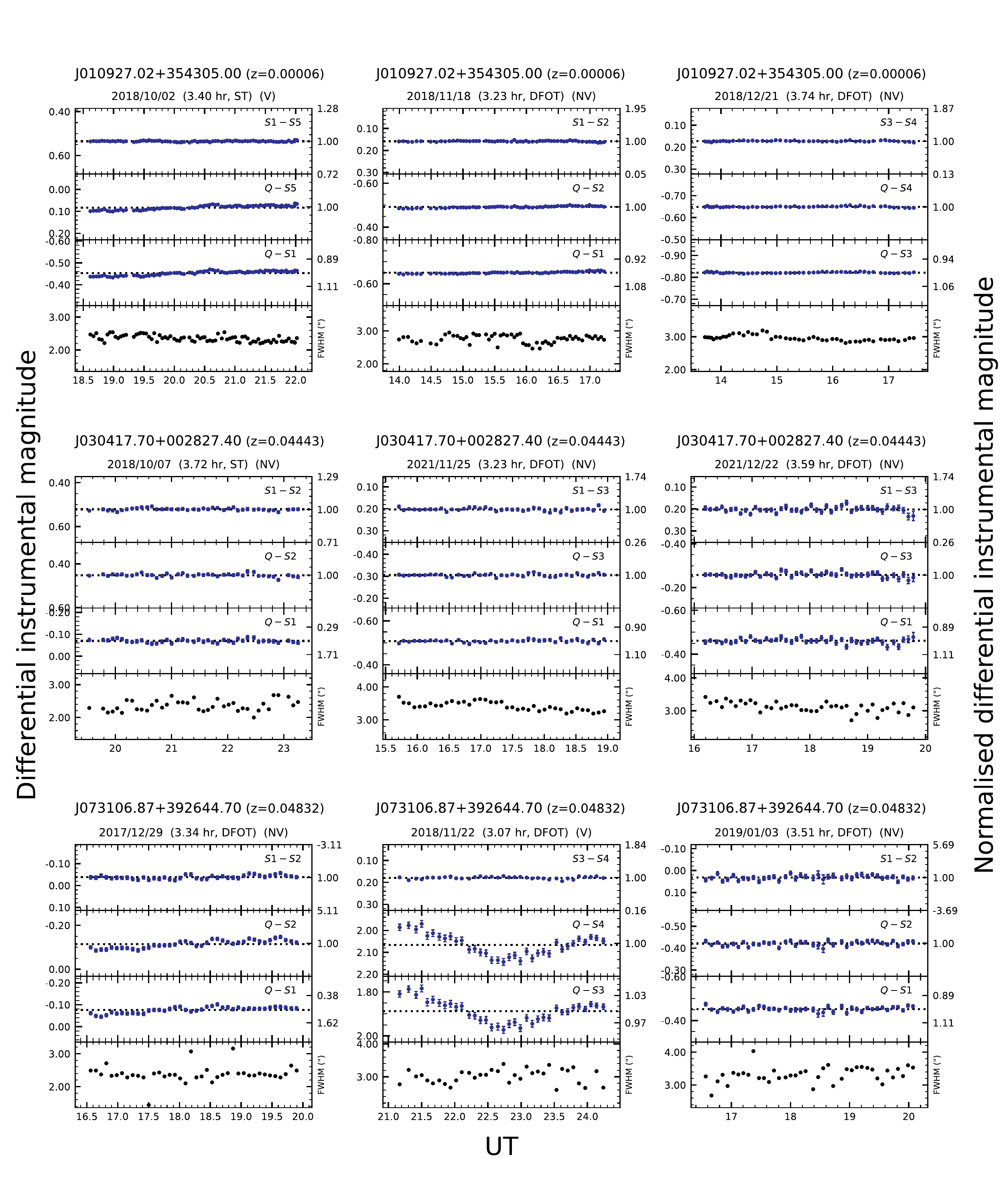}\\
  \vspace{-0.1in}
  \caption{Differential light curves (DLCs) for our sample of 12 low-mass, radio/X-ray detected AGN. The name of the LMAGN, its redshift and the date $\&$ duration of
 the monitoring session are given at the top of each panel, together with the INOV status for the session. In each panel, the top panel displays the DLC for the pair of comparison stars. The subsequent two panels show the DLCs of the AGN relative to the two comparison stars, as defined in the labels on the right side. The bottom panel shows the variation of the seeing disk (FWHM), i.e., `point spread function' (PSF) through the monitoring session. The labels on the right side of each panel show the  differential instrumental magnitude normalised by  mean value, as marked by the dotted horizontal line.}
\label{fig1} 
\end{figure*}

\begin{figure*}
  \includegraphics[width=1.09\textwidth,height=0.95\textheight]{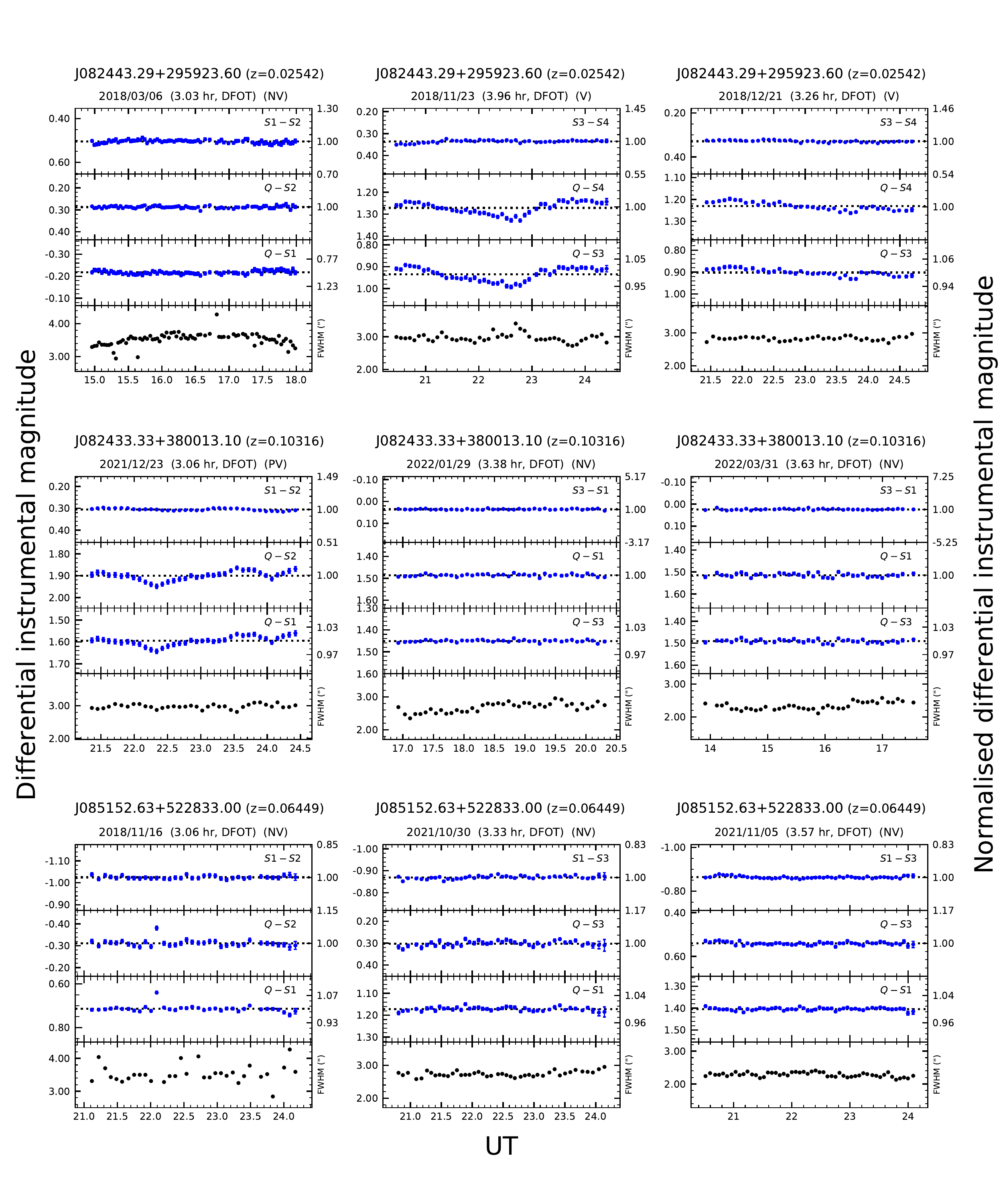}\\

  \vspace{-0.1in}
  \caption{Same as Fig.~\ref{fig1}.}
\label{fig2} 
\end{figure*}

\begin{figure*}
  
   \includegraphics[width=1.09\textwidth,height=0.95\textheight]{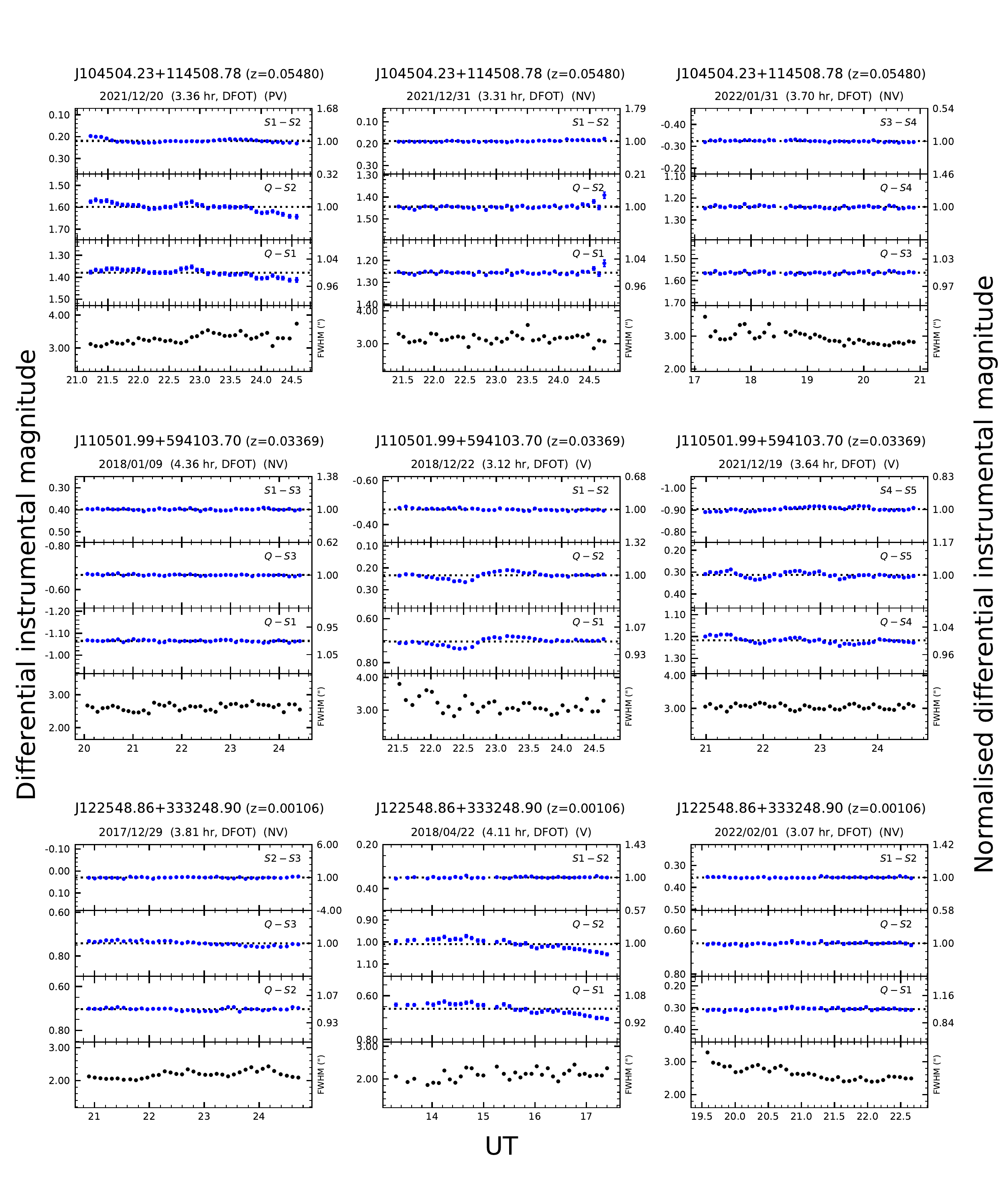}\\

  \vspace{-0.1in}
  \caption{Same as Fig.~\ref{fig1}.}
\label{fig3} 
\end{figure*}

\begin{figure*}
 
  \includegraphics[width=1.09\textwidth,height=0.95\textheight]{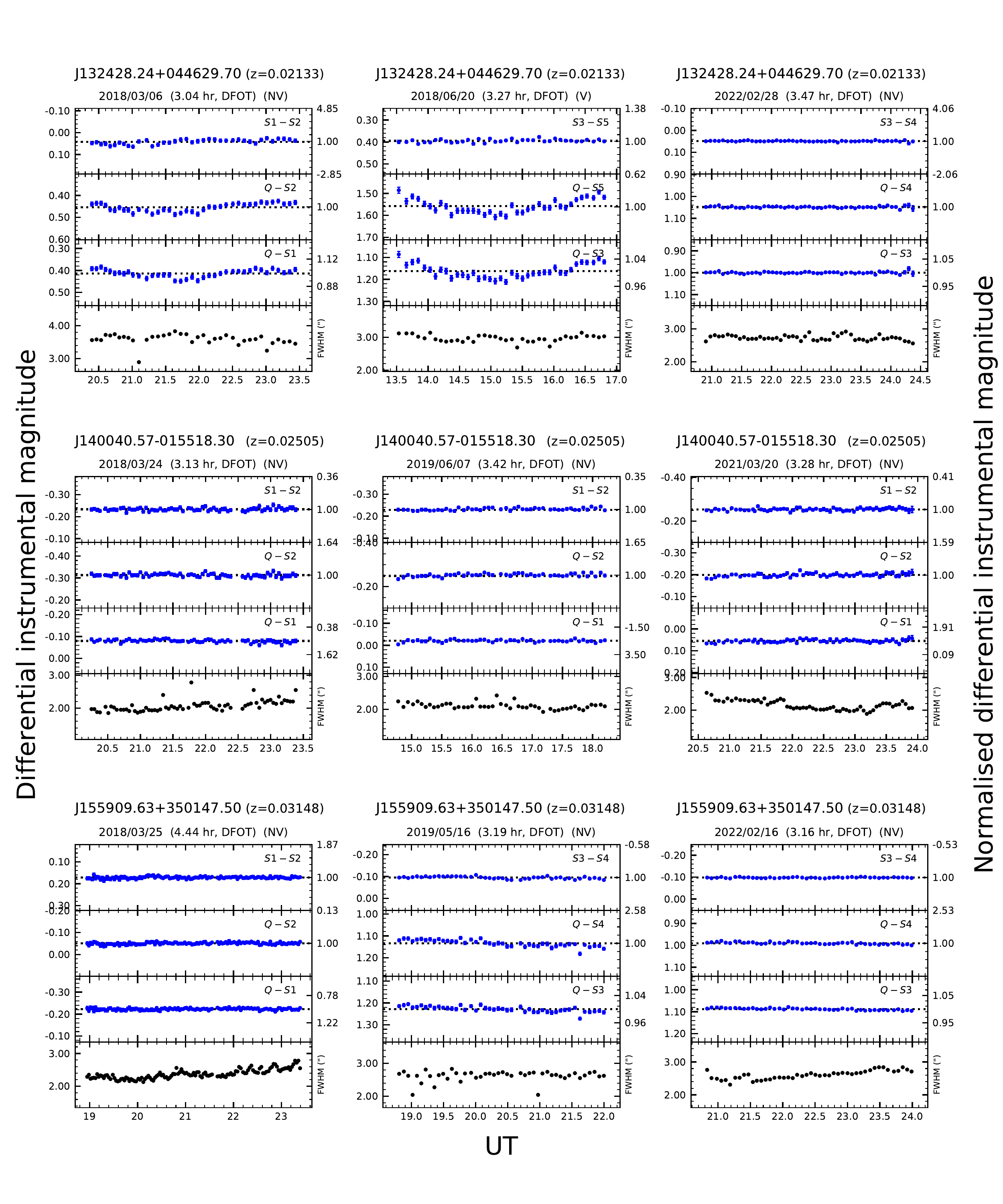}\\
  
  \vspace{-0.1in}
  \caption{Same as Fig.~\ref{fig1}.}

\label{fig4}
\end{figure*}

\clearpage

\floatsetup[table]{capposition=top}
\begin{table*}
 \caption{Basic parameters of selected pool of comparison stars.
  \label{tab:comparison_color}}
\scriptsize
\begin{tabular}{ccc ccc r r}\\
\hline

{Target LMAGN and } &   Date(s) of monitoring       &   {R.A.(J2000)} & {Dec.(J2000)}             & {\it g} & {\it r} & \hspace{0.15 cm} {\it g-r}  & { $\Delta(g-r)$} \\
the comparison stars
                  &  yyyy/mm/dd   &   (hh:mm:ss)       &($^\circ$: $^\prime$: $^{\prime\prime}$)   & (mag)   & (mag)   & (mag)  & Q-S(mag)     \\
{(1)}      & {(2)}        & {(3)}           & {(4)}                              & {(5)}   & {(6)}   &  \hspace{0.15 cm}{(7)}  &(8)   \\
\hline
\multicolumn{7}{l}{}\\

J010927.02$+$354305.00$^{*}$  & 2018/10/02, 2018/11/18, 2018/12/21 & 01:09:27.02&$+$35:43:05.00 &16.20  &17.50   & $-$1.30  & - \\
                                                       
S1            & 2018/10/02, 2018/11/18 & 01:09:18.06&$+$35:44:31.44 &14.70  &16.00   & $-$1.30 & 0 \\
                                                      
S2            & 2018/11/18  & 01:09:28.45&$+$35:44:17.91 &16.20  &17.50   & $-$1.30  & 0 \\
S3            & 2018/12/21 & 01:09:54.63&$+$35:49:29.15 &13.70  &13.60   &0.10   & $-$1.4\\
                                                      
S4            & 2018/12/21  & 01:08:53.79&$+$35:48:46.29 &14.10  &13.20   &0.90 & $-$2.2  \\

S5            & 2018/10/02  & 01:09:14.07&$+$35:42:54.04 &14.40  &16.20   & $-$1.80   & 0.5 \\\\


J030417.70$+$002827.40 & 2018/10/07, 2021/11/25, 2021/12/22 & 03:04:17.70&$+$00:28:27.40 & 15.83 &  15.19 & 0.64 & -  \\
S1           & 2018/10/07, 2021/11/25, 2021/12/22 & 03:04:04.09&$+$00:26:41.72 & 16.47 &  16.10 & 0.37 & 0.27 \\
S2           & 2018/10/07             & 03:04:03.26&$+$00:28:17.96 & 16.11 &  15.59 & 0.52   & 0.12 \\
S3           & 2021/11/25,  2021/12/22                        & 03:04:45.25&$+$00:32:57.30 & 16.59 &  15.95 & 0.65 & $-$0.01 \\\\

J073106.87$+$392644.70 & 2017/12/29, 2018/11/22, 2019/01/03 & 07:31:06.87&$+$39:26:44.70 & 15.89 &  15.20 & 0.69 & -  \\
S1           & 2017/12/29, 2019/01/03             & 07:31:19.29&$+$39:25:59.29 & 16.67 &  16.31 & 0.36   & 0.23 \\
S2           & 2017/12/29, 2019/01/03             & 07:31:18.79&$+$39:26:21.67 & 16.69 &  16.36 & 0.33   & 0.36\\
S3           & 2018/11/22                         & 07:31:25.07&$+$39:19:01.50 & 15.07 &  14.68 & 0.39   & 0.30\\
S4           & 2018/11/22                         & 07:31:21.90&$+$39:24:45.16 & 15.67 &  14.49 & 1.18   & $-$0.49\\\\

J082443.29$+$295923.60 & 2018/03/06, 2018/11/23, 2018/12/21   & 08:24:43.29&$+$29:59:23.60 & 15.92 &  15.38 & 0.54 & - \\
                                                       
                                                       
S1           &2018/03/06                 & 08:24:51.09&$+$30:04:16.55 & 16.08 &  16.66 & $-$0.58   & 2.11 \\
S2           &2018/03/06                & 08:24:52.19&$+$30:04:21.03 & 15.28 &  15.10 & 0.18   & 0. 36\\            
S3           &2018/11/23, 2018/12/21                            & 08:25:05.52&$+$30:05:24.30 & 15.02 &  14.23 & 0.79  & $-$0.25 \\
S4           &2018/11/23, 2018/12/21                            & 08:24:34.58&$+$29:58:31.40 & 14.42 &  13.99 & 0.43  &  0.11\\\\

                                                      
                                                      

J082433.33$+$380013.10 & 2021/12/23, 2022/01/29, 2022/03/31    & 08:24:33.33&$+$38:00:13.10 & 16.63 &  16.21 & 0.42  & - \\
                                 
S1           & 2021/12/23, 2022/01/29, 2022/03/31    & 08:23:51.80&$+$38:05:46.68 & 15.29 &  14.86 & 0.43  &  $-$0.01\\
                                 
S2           & 2021/12/23     & 08:24:11.84&$+$37:57:18.77 & 14.90 &  14.56 & 0.34   & 0.08 \\
S3           & 2022/01/29, 2022/03/31    & 08:24:11.06&$+$38:07:37.49 & 15.42 &  14.89 & 0.53   & $-$0.11 \\\\

 J085152.63$+$522833.00 & 2018/11/16, 2021/10/30, 2021/11/05  & 08:51:52.63&$+$52:28:33.00 & 16.45 &  15.92 & 0.53 & -  \\
                                                    
 S1 	     & 2018/11/16, 2021/10/30, 2021/11/05  & 08:52:09.74&$+$52:22:34.87 & 15.64 &  15.26 & 0.38 &  0.15 \\
                                                    
 S2 	     & 2018/11/16              & 08:51:57.24&$+$52:21:18.26 & 16.81 &  16.26 & 0.55  & $-$0.02 \\
 S3          & 2021/10/30, 2021/11/05    & 08:52:18.85&$+$52:23:34.33 & 17.21 &  16.17 & 1.04& $-$0.51 \\\\

 J104504.23$+$114508.78 &  2021/12/20, 2021/12/31, 2022/01/31 & 10:45:04.23 &$+$11:50:08.78 & 16.56 &  16.01 & 0.55  & - \\
 S1 	      &  2021/12/20, 2021/12/31             & 10:45:24.29 &$+$11:40:25.50 & 15.61 &  15.06 & 0.55   & 0 \\
 S2 	      &  2021/12/20, 2021/12/31             & 10:44:45.49 &$+$11:30:00.70 & 15.31 &  14.86 & 0.45   & 0.10\\
 S3           &  2022/01/31             & 10:45:36.36 &$+$11:49:50.51 & 15.05 &  14.70 & 0.35  & 0.20 \\
 S4           &  2022/01/31             & 10:45:17.85 &$+$11:49:08.63 & 15.64 &  15.00 & 0.64 & $-$0.09\\\\
 
J110501.99$+$594103.70 & 2018/01/09, 2018/12/22, 2021/12/19 & 11:05:01.99&$+$59:41:03.70 & 15.15 &  14.58 & 0.57   &- \\
                                                     
S1           & 2018/01/09, 2018/12/22 & 11:05:45.37&$+$59:39:21.45 & 17.36 &  15.93 & 1.43   & $-$0.86\\
                                                     
S2           & 2018/12/22             & 11:04:57.19&$+$59:43:39.64 & 14.84 &  14.34 & 0.50   & 0. 07\\
                                                      
S3           & 2018/01/09                         & 11:05:50.64&$+$59:35:40.36 & 16.78 &  15.55 & 1.23  & $-$0.66 \\

S4           & 2021/12/19                         & 11:05:56.61&$+$59:33:37.01 & 14.47 &  13.95 & 0.52   & 0.05 \\
S5           & 2021/12/19                         & 11:05:18.78&$+$59:45:05.71 & 15.45 &  14.82 & 0.63   & $-$0.05 \\\\

J122548.86$+$333248.90 & 2017/12/29, 2018/04/22, 2022/02/01 & 12:25:48.86&$+$33:32:48.90 & 14.24 &  13.74 & 0.50  & - \\
S1 	     & 2018/04/22, 2022/02/01 & 12:25:44.88&$+$33:38:37.50 & 15.97 &  15.59 & 0.38   & 0.12\\
S2 	     & 2017/12/29, 2018/04/22, 2022/02/01 & 12:25:20.51&$+$33:33:00.88 & 16.49 &  15.32 & 1.17  & $-$0.67 \\
S3           & 2017/12/29                         & 12:26:22.16&$+$33:33:13.62 & 15.60 &  15.21 &0.39  & 0.11\\\\

J132428.24$+$044629.70 & 2018/03/06, 2018/06/20, 2022/02/28 & 13:24:28.24&$+$04:46:29.70 & 16.09 &  15.33 & 0.76  & - \\
S1 	     & 2018/03/06                         & 13:24:51.05&$+$04:51:26.47 & 15.69 &  15.26 & 0.43   &  0.33\\
S2 	     & 2018/03/06             & 13:24:45.05&$+$04:43:27.84 & 16.50 &  15.24 & 1.26   & $-$0.50\\
S3           & 2018/06/20, 2022/02/28 & 13:24:40.58&$+$04:42:56.05 & 15.11 &  14.58 & 0.53  & 0.23\\
S4           & 2022/02/28                         & 13:24:45.94&$+$04:52:43.36 & 15.67 &  14.60 & 1.07 &  $-$0.31\\
S5           & 2018/06/20                         & 13:24:49.31&$+$04:38:44.09 & 14.78 &  14.20 & 0.58  & 0.18\\\\

J140040.57$-$015518.30 &2018/03/24, 2019/06/07, 2021/03/20  & 14:00:40.57&$-$01:55:18.30 & 15.91 &  15.40  & 0.51  & - \\        
S1 	     &2018/03/24, 2019/06/07, 2021/03/20  & 14:00:21.27&$-$01:58:08.63 & 16.16 &  15.54  & 0.62   & $-$0.11\\
S2 	     &2018/03/24, 2019/06/07, 2021/03/20  & 14:00:23.19&$-$02:01:13.88 & 16.41 &  15.81  & 0.60   & $-$0.0.9\\\\

J155909.63$+$350147.50 & 2018/03/25, 2019/05/16, 2022/02/16  & 15:59:09.63&$+$35:01:47.50 & 14.61 &  14.03 & 0.58  & -  \\
                                                       
S1           & 2018/03/25              & 15:59:28.77&$+$35:07:41.25 & 15.70 &  15.09 & 0.61   & $-$0.03 \\
                                                      
S2           & 2018/03/25              & 15:59:18.53&$+$34:54:43.73 & 15.54 &  14.82 & 0.72   & $-$0.14\\
                                                       
S3           & 2019/05/16, 2022/02/16  & 15:59:35.02&$+$35:00:42.00 & 15.48 &  16.08 & $-$0.60   & 1.18\\
S4           & 2019/05/16, 2022/02/16  & 15:59:39.51&$+$35:01:58.33 & 15.11 &  14.09 & 1.02   & $-$0.44\\\\
\hline\\
\multicolumn{7}{l}{Due to unavailability of SDSS (g-r) color for J010927.02$+$354305.00 marked with `$^{*}$', its (B-R) colour has been taken from}\\ 
\multicolumn{7}{l}{USNO-A2.0 catalogue (\citealt{Monet1998AAS...19312003M}).}
\end{tabular}

\end{table*}

\floatsetup[table]{capposition=top}
\begin{table*}
 \begin{minipage}[10]{500mm}
  
\caption{Result of the statistical test for detecting INOV in the DLCs of the 12 LMAGNs (taking $\eta=1.54$, see Appendix A). }
\label{tab2}
\hspace{-0.5in}
\begin{tabular}{@{}ccccccccccccc@{}}
  \hline \hline
\multicolumn{1}{c}{LMAGN}
&\multicolumn{1}{c}{Date}
&{N}
&{T} 
&\multicolumn{3}{c}{F-test}
 &\multicolumn{2}{c}{INOV}
 
 &\multicolumn{1}{c}{$\sqrt { \eta^2\langle \sigma^2_{i,err} \rangle}$}
 & $\overline\psi$ &\\
  (SDSS name)       & yyyy/mm/dd & &{(hr)} &{$F_{1}^{\eta}$},{$F_{2}^{\eta}$}&{$F_{c}(0.95)$}&{$F_{c}(0.99)$}&\multicolumn{2}{c}{status$^{a}$} && \%&\\
 (1)     &   (2)    &(3)&(4)  & (5)                         &(6)       &(7)           &(8)            &(9)     &(10)     & (11)     \\
\hline\hline\\

J010927.02+354305.00 & 2018/10/02 & 77   &  3.40&     8.29, 8.25  &  1.46&  1.71&     V        ,     V     & V        & 0.002 &  3.67       \\
J010927.02+354305.00 & 2018/11/18 &  59 &  3.23&      1.81, 1.54   &  1.55&  1.86&     PV        ,    NV    & NV       & 0.003 &  ---  \\
J010927.02+354305.00 & 2018/12/21 &  49 &  3.74&      1.01, 1.05   &  1.62&  1.98&     NV        ,    NV    & NV       & 0.003 &  --- \\\\

J030417.70+002827.40 & 2018/10/07 &  40  &  3.72&      0.46, 0.61  &  1.70&  2.14&     NV       ,     NV    & NV       & 0.009 &  ---    \\
J030417.70+002827.40 & 2021/11/25 &  39  &  3.23&      0.49, 0.55  &  1.72&  2.16&     NV       ,     NV     & NV      & 0.009 &  ---     \\
J030417.70+002827.40 & 2021/12/22 &  42  &  3.59&      0.53, 0.53  &  1.68&  2.09&     NV       ,     NV     & NV      & 0.016 &  ---     \\\\

J073106.87+392644.70 & 2017/12/29 & 40   &  3.34&      1.53, 2.44  &  1.70&  2.14&     NV       ,      V    & NV      & 0.011 &  ---     \\
J073106.87+392644.70 & 2018/11/22 & 35   &  3.07&      5.11, 4.93  &  1.77&  2.26&     V         ,      V     & V       & 0.005 & 17.88  \\
J073106.87+392644.70 & 2019/01/03 &  40  &  3.51&      0.56, 0.62  &  1.70&  2.14&     NV       ,     NV    & NV       & 0.015 &  ---      \\\\
J082443.29+295923.60 & 2018/03/06 & 78   &  3.03&      0.43, 0.44  &  1.46&  1.71&      NV       ,      NV    & NV      & 0.010 & --- \\
J082443.29+295923.60 & 2018/11/23 & 46   &  3.96&      6.61, 6.14  &  1.64&  2.02&      V       ,      V    & V      & 0.006 & 9.82 \\
J082443.29+295923.60 & 2018/12/21 & 38   &  3.26&      2.85, 4.36  &  1.73&  2.18&      V       ,      V    & V      & 0.005 &  5.97  \\\\

J082433.33+380013.10 & 2021/12/23 & 36   &  3.06&      1.97, 2.07  &  1.76&  2.23&      PV       ,     PV    & PV      & 0.004 &  8.12 \\
J082433.33+380013.10 & 2022/01/29 & 40   &  3.38&      0.63, 0.66  &  1.70&  2.14&      NV       ,     NV    & NV      & 0.003 &  ---  \\
J082433.33+380013.10 & 2022/03/31 & 40   &  3.63&      0.66, 0.77  &  1.70&  2.14&      NV       ,     NV    & NV      & 0.004 &  ---  \\\\

J085152.63+522833.00 & 2018/11/16 &   35 &  3.06&      2.53, 1.14  &  1.77&  2.26&      V       ,     NV    & NV     & 0.012 &  ---  \\
J085152.63+522833.00 & 2021/10/30 &   43 &  3.33&      0.56, 0.73  &  1.67&  2.08&     NV       ,     NV    & NV     & 0.008 &  ---  \\
J085152.63+522833.00 & 2021/11/05 &  51  &  3.57&      0.55, 0.51  &  1.60&  1.95&     NV       ,     NV    & NV     & 0.006 &  ---  \\\\

J104504.23+114508.78 & 2021/12/20 & 39   &  3.36&      1.86, 2.85  &  1.72&  2.16&      PV       ,      V    &  PV      & 0.005 &  6.66  \\
J104504.23+114508.78 & 2021/12/31 & 39   &  3.31&      1.21, 1.80  &  1.72&  2.16&      NV       ,     PV    & NV      & 0.004 &  ---  \\
J104504.23+114508.78 & 2022/01/31 & 42   &  3.70&      0.57, 0.52  &  1.68&  2.09&      NV       ,     NV    & NV      & 0.003 &  ---  \\\\

J110501.99+594103.70 & 2018/01/09 &   42&4.36  &      0.82, 0.52  &  1.68&  2.09&     NV       ,     NV    & NV     & 0.006 &  ---     \\
J110501.99+594103.70 & 2018/12/22 &    36&  3.12&       5.64, 3.45 &  1.76&  2.23&      V       ,     V     & V     & 0.005 &  5.56 \\
J110501.99+594103.70 & 2021/12/19 &    43&  3.64&       4.31, 2.62 &  1.67&  2.08&      V       ,     V     & V     & 0.005 &  4.85 \\\\

J122548.86+333248.90 & 2017/12/29 &     37 &3.81&      1.46, 4.28  &  1.74&  2.21&      NV       ,     V    &  NV    & 0.003 &  ---     \\
J122548.86+333248.90 & 2018/04/22 &      35&4.11&      5.21, 5.45  &  1.77&  2.26&      V       ,     V     & V    & 0.005 &  7.98  \\
J122548.86+333248.90 & 2022/02/01 &      37&3.07&      0.34, 0.32  &  1.74&  2.21&     NV       ,     NV    & NV    & 0.004 &  ---      \\\\

J132428.24+044629.70 & 2018/03/06 &     38 &3.04 &     1.59, 1.88 &  1.73&  2.18&    NV       ,    PV     & NV    & 0.011 & ---   \\
J132428.24+044629.70 & 2018/06/20 &     38 &3.27&        3.43, 3.66&  1.73&  2.18&      V       ,     V     & V    & 0.008 & 12.17  \\
J132428.24+044629.70 & 2022/02/28 &     50 &3.47&        0.41, 0.48&  1.61&  1.96&     NV       ,     NV    & NV     & 0.003 &  --- \\\\

J140040.57-015518.30 & 2018/03/24 &      68&3.13&        0.47, 0.47&  1.50&  1.78&     NV       ,     NV    & NV   & 0.011 &  ---    \\
J140040.57-015518.30 & 2019/06/07 &      48&3.42&        0.45, 0.59&  1.62&  1.99&     NV       ,     NV    & NV   & 0.008 &  ---     \\
J140040.57-015518.30 & 2021/03/20 &      61&3.28&        0.60, 0.63&  1.53&  1.84&     NV       ,     NV    & NV    & 0.009 &  ---    \\\\

J155909.63+350147.50 & 2018/03/25 &     125&4.44&        0.34, 0.45&  1.35&  1.52&     NV       ,     NV    & NV     & 0.007 &  ---     \\
J155909.63+350147.50 & 2019/05/16 &      45&3.19&        1.49, 2.17&  1.65&  2.04&     NV       ,     V    & NV     & 0.004 &  ---     \\
J155909.63+350147.50 & 2022/02/16 &      46&3.16&        0.59, 0.67&  1.64&  2.02&     NV       ,     NV    & NV     & 0.002 &  ---     \\\\

\hline
\multicolumn{11}{l}{$^a$ V=variable, i.e., confidence
       $\ge 0.99$; PV = probable variable ($0.95-0.99)$; NV
       = non-variable ($< 0.95$).}\\
\multicolumn{11}{l}{Variability status identifiers (col. 8), based on AGN-star1 and AGN-star2 DLCs are separated by a comma.}
\end{tabular}
 \end{minipage}
\end{table*}



 
 \clearpage

\appendix

\section{Statistical analysis of the differential light-curves (DLCs)}

\label{stat_ana}
 For each AGN, differential light curves (DLCs) for a session were determined relative to two (steady) comparison stars. The selected two comparison stars for each 
 session and the labels of the 3 DLCs involving them are shown in the Figs. \ref{fig1}-\ref{fig4} on the right side. Recall that in several independent studies, it has 
 been shown that the photometric errors returned by DAOPHOT are underestimated \citep{Gopal-Krishna1995MNRAS.274..701G,Stalin2004JApA...25....1S,Bachev2005MNRAS.358..774B} by a factor $\eta$ = 1.54, as estimated in an extensive study by \citet{Goyal2013JApA...34..273G}, based on INOV data acquired in 262 sessions, and the same value of 
 $\eta$ has been adopted here. To check for INOV in the DLCs, we have employed the F$-\eta$ test (\citealp{Diego2010AJ....139.1269D}; \citealp{Villforth2010ApJ...723..737V}; \citealp{2012A&A...544A..37G}) as also employed in \citet{Goyal2013MNRAS.435.1300G} in their homogeneous analysis of DLCs from 262 intranight sessions covering 6 prominent classes of luminous AGN. The $F$-values for the two LMAGN DLCs for a session are:
\begin{equation} 
\label{eq.ftest2}
F_{1}^{\eta} = \frac{Var(q-s1)}{ \eta^2 \sum_\mathbf{i=1}^{N}\sigma^2_{i,err}(q-s1)/N}  \\
,\hspace{0.1cm} F_{2}^{\eta} = \frac{Var(q-s2)}
{ \eta^2 \sum_\mathbf{i=1}^{N}\sigma^2_{i,err}(q-s2)/N}  \\
\end{equation}\\
where $Var(q-s1)$ and $Var(q-s2)$ are the variances of the DLCs of the target LMAGN, relative to the two chosen comparison stars, and $\sigma_{i,err}(q-s1)$ and 
$\sigma_{i,err}(q-s2)$ represent the rms error returned by DAOPHOT on the $i^{th}$ data point in the DLCs of the target LMAGN, relative to the two comparison stars.
N is number of data points in the DLCs and the scaling factor  ${\eta= 1.54}$, as mentioned above.
Table \ref{tab2} lists N and the computed values of $F_{1}^{\eta}$ and $F_{2}^{\eta}$ for the two DLCs of the target LMAGN for each session.\par
The critical values of $F $ ($= F_{c}^{\alpha}$) for $\alpha = $ 0.05, 0.01 correspond to confidence levels of 95\% and 99\%, respectively. For each session, these two computed values are listed in columns 6 $\&$ 7 in Table \ref{tab2} and they are compared with the $F-$values computed for the two DLCs of the LMAGN using Eq. (A1), namely, 
$F_{1,2}^{\eta}$ (Column 5 in Table \ref{tab2}). If the computed $F-$value for a DLC of the target LMAGN exceeds the critical value $F_{c}$ for that session, the null hypothesis (i.e., no variability) is discarded. For a computed $F-$value $\ge$ $F_{c}$(0.99), the DLC of the target LMAGN is classified as `variable' (V). The designation is `probable variable' (PV) if the computed $F-$value falls between $F_{c}$(0.95) and $F_{c}$(0.99), and `non-variable'  (NV) if the $F-$value is less than $F_{c}$(0.95). Note that the designation for a given session, as given in column 9 of the Table \ref{tab2}, is `V' only if both DLCs of the target LMAGN belong to the `V'  category and `NV' if even one of the two DLCs is of the `NV' type. The remaining sessions have been designated `probable variable' (PV). Column 10 of Table \ref{tab2} lists for each session the `Photometric Noise Parameter' (PNP) = {$\sqrt { \eta^2\langle \sigma^2_{i,err} \rangle}$ }, where ${\eta=1.54}$, as mentioned above.

The variability amplitude ($\psi$) for a DLC is defined as \citep{Heidt1996A&A...305...42H}: $\psi= \sqrt{({A_{max}}-{A_{min}})^2-2\sigma^2}$\\
Here $A_{max}$ and $A_{min}$ are the maximum and minimum values in the LMAGN-star DLC and $\sigma^2=\eta^2<\sigma^2_{q-s}>$, where, $¡´\sigma^2_{q-s}¡µ$ is the mean 
square rms error for the data points in the DLC and the error underestimation factor ${\eta=1.54}$. Column 11 of the Table \ref{tab2} gives the mean value of $\psi$ 
for a session, i.e., the average of the $\psi$ values estimated for the two DLCs of the target LMAGN.

\label{lastpage}
\end{document}